Best Practices

# Crowdfunding Astronomy Outreach Projects: Lessons learned from the UNAWE crowdfunding campaign

**Abi J. Ashton**
Leiden University/UNAWE, The Netherlands
ashton@strw.leidenuniv.nl

**Pedro Russo**
Leiden University/UNAWE, The Netherlands
russo@strw.leidenuniv.nl

**Thilina Heenatigala**
Leiden University/UNAWE, The Netherlands
heenatigala@strw.leidenuniv.nl

## Summary

In recent years, crowdfunding has become a popular method of funding new technology or entertainment products, or artistic projects. The idea is that people or projects ask for many small donations from individuals who support the proposed work, rather than a large amount from a single source. Crowdfunding is usually done via an online portal or platform which handles the financial transactions involved. The Universe Awareness (UNAWE) programme decided to undertake a Kickstarter[1] crowdfunding campaign centring on the resource Universe in a Box[2]. In this article we present the lessons learned and best practices from that campaign.

## Why crowdfunding?

For those who have not been involved in a crowd-funded project, it can look like an easy way to raise money. An idea is presented on a platform, where many other ideas are presented, and in an ideal world lots of people put money into it and the project becomes a success. This is not quite how it works.

The idea that it is easy is perhaps the result of the simple nature of the language and the videos used on these sites. Grant proposals are filled with highly technical language, whereas a crowdfunding page has to appeal to, and be understood by, a wider audience.

Crowdfunding is a lot of work. By its nature it requires the mobility and participation of a large group of stakeholders. Many people have to be persuaded that the idea is a good one, not just one large potential investor. On the plus side the effort that goes into selling an idea for crowdfunding brings with it excellent publicity and awareness in addition to a source of revenue.

## Crowdfunding platforms

Crowdfunding platforms create the necessary organisational systems and conditions for crowdfunding projects to take place. By replacing traditional intermediaries between supply and demand — such as traditional record companies and venture capitalists — platforms such as Kickstarter link new artists, designers and project initiators with supporters who believe in the project enough to invest financial support.

With so many platforms to choose from it is important to research thoroughly before choosing one. Table 1 is only a selection of those available. This article mainly looks at reward- and donation-based platforms as these are considered the most interesting for astronomy outreach projects, but other platforms can also be equity or credit-based.

| Name | Focus | Link | Funding model |
|---|---|---|---|
| Kickstarter | Creative projects | www.kickstarter.com | AoN |
| Indiegogo | Allows a broad range of projects | www.indiegogo.com | KiA |
| Rockethub | Broad range of projects | www.rockethub.com | KiA |
| 1%Club | Development projects | www.onepercentclub.com/en/ | KiA |
| Global Giving | Donation/charity | www.globalgiving.com | KiA |
| Crowdcube | Start-ups | www.crowdcube.com | AoN |

*Table 1.* Overview of the most popular crowdfunding platforms.

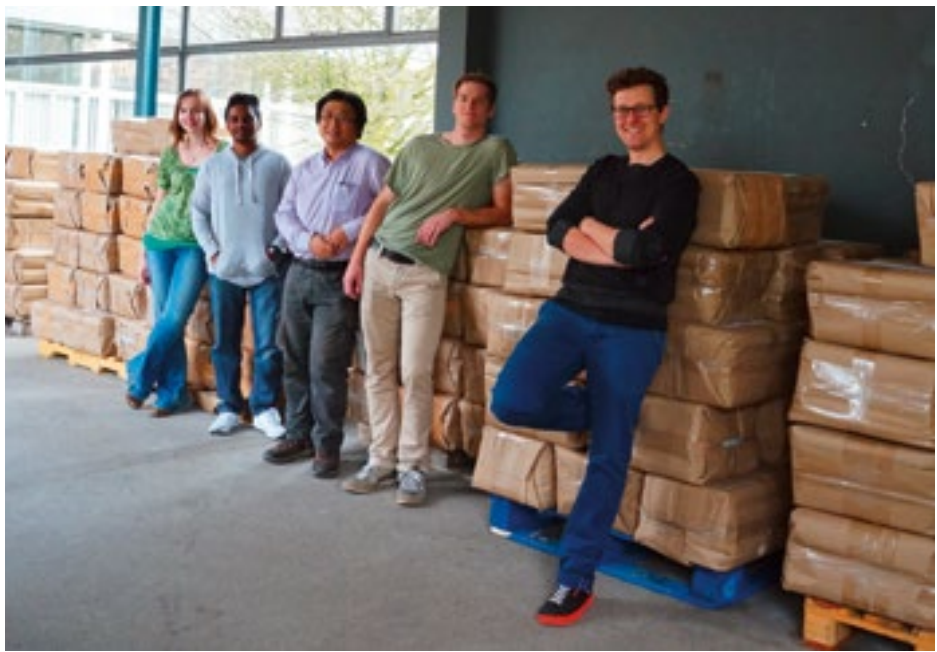

*Figure 1.* Universe in a Box kits packed and ready for shipping around the world. Credit: C. Leung (UNAWE)

Of particular note when choosing a platform is the funding model used, which will be either All or Nothing (AoN) or Keep it All (KiA). Under the AoN system, money is only collected from contributors if a pre-determined target has been pledged. If the goal is not met, no money is collected. KiA means that all of the funds pledged (minus fees) are handed over to the project regardless of whether there are sufficient funds to meet objectives.

### The UNAWE campaign

The UNAWE programme took the decision to undertake a Kickstarter crowdfunding campaign centring on the resource, Universe in a Box. Universe in a Box is a physical science education kit to assist teachers and educators in bringing astronomy and space science to four-to ten-year-old children around the world.

### Goal

The goal of the project was to raise 15 000 euros to cover the cost of distributing 160 Universe in a Box kits around the world to underprivileged communities and producing online training videos (Figure 1). Calculated within this was the amount it would cost to ship campaign rewards and the fees which Kickstarter takes from the total.

### Budget

The budget for our campaign was 800 euros to produce a campaign video[3] (Figure 2) in addition to three months full-time-equivalent personnel for before, during and after the campaign.

### Campaign overview

UNAWE has an extensive international network which gives a strong foundation from which to launch a crowdfunding campaign, but thorough preparation is also important for success. Before the launch date the equivalent of almost one month of full-time work was spent preparing text and visual content for the Kickstarter page and coming up with a plan of action for during the campaign itself. The Kickstarter campaign page[4] (Figure 3) was talked about and shared before launch to enable a hype to build.

The campaign was promoted using email and social media, as well as through contacts with several journalists and conversations with personal friends and other contacts. The team contributed to several blogs about the campaign and other people also blogged about the project.

Updates from UNAWE were posted on Twitter and Facebook accounts at least twice daily, trying to use images once a day and attempting to make sure that the content was diverse. For example, on one day a particular reward would be promoted, and on the next the number of pledgers would be highlighted. The UNAWE Facebook page is liked by 2343 people and the Twitter page is followed by 4627 people at the time of writing.

The website Peerreach[5] was used to identify high-impact Twitter users, particularly within the relevant fields of science and education. These people were then sent personalised messages with a link to the campaign. We aimed to send 20 of these every weekday to different people.

Emails were sent out to various groups of potentially interested people throughout the campaign and to the UNAWE mailing lists at the beginning, middle and end of the campaign.

The UNAWE international mailing list has 1771 subscribers and the UNAWE Dutch mailing list has 844 subscribers. Press releases were prepared and there was media coverage through a variety of outlets.

### Results

At the end of the 31-day campaign 17 037 euros had been pledged, bringing the project above the 15 000-euro goal (Figure 4). After fees had been deducted

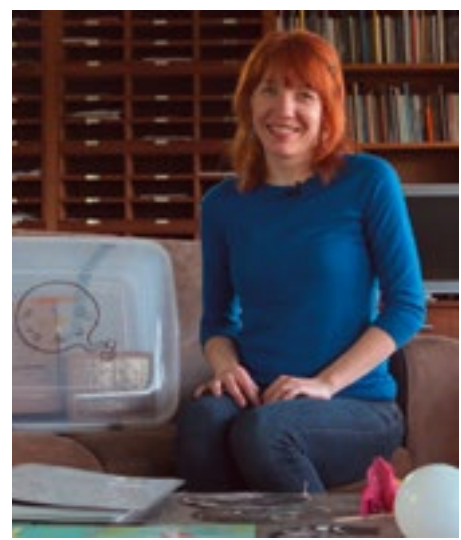

*Figure 2.* Screenshot from the presentation film for the campaign with the astronomer Monica Turner. Credit: van Schadewijk. R/UNAWE

and a couple of pledges dropped, and a sum of 15 463.56 euros was received from Kickstarter.

The campaign had 235 backers with an average backing of 72.50 euros, was shared 1395 times on Facebook, and the video was viewed 2664 times with 45.14% of plays completed.

Table 2 is a direct product from the Kickstarter page and shows where pledgers arrived at the Kickstarter page from. Most were direct traffic, but a significant proportion of pledgers arrived at the site from the Twitter and Facebook links.

Blog articles and a widget embedded on the UNAWE website also contributed a reasonable amount to the campaign.

## Follow-up

Once the campaign was complete, there were three main obligations: shipping the rewards to project backers, fulfilling the targets set out by the campaign, and continuing to build and strengthen the community.

## Lessons learned and recommendations

Much was learnt from this foray into the world of crowdfunding and here we will look at some of these lessons, and then summarise recommendations for any other astronomy outreach projects looking to raise funds this way.

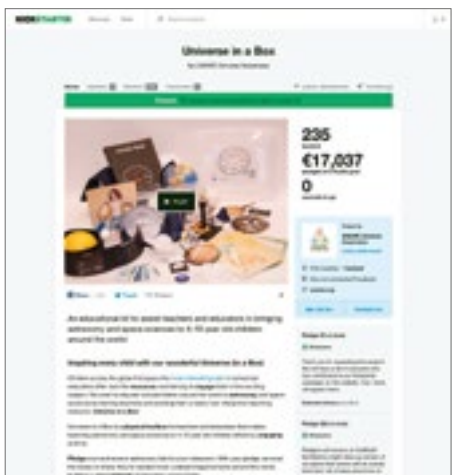

*Figure 3.* Kickstarter Universe in a Box campaign page.

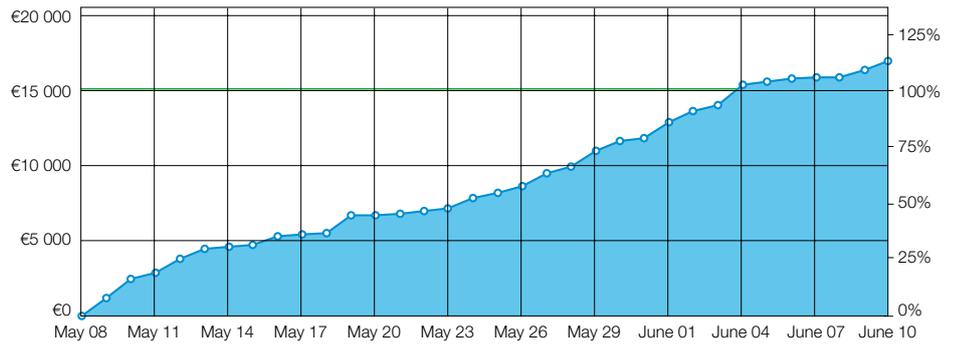

*Figure 4.* Evolution of the donations during the campaign period

### 1. Don't do it only for the funds

A crowdfunding campaign should also aim to build awareness and a community, as well as raise funds. Community and awareness-building should be as much a part of your goal when running a crowdfunding project as the financial reward.

### 2. Be prepared

Work on a promotion campaign beforehand and try to have some confirmed pledges beforehand — some say 50%, we had around 25%. Invest in a good video, which should be professional, but not overdone — our video had 2495 views. You should also have a clear plan of how to carry out your campaign if it is successful.

### 3. Have a clear budget

You need a clear budget for the project, not just a rough estimate. Backers want to know what you will do with their money and you should be prepared to answer any questions they have, like what percentage of funds will be spent on travel or distribution. Don't forget to include any fees for the platform you are using and the cost of reward fulfilment for your backers — how much will it cost to actually send those rewards around the world?

### 4. Be very active on social media

**Twitter: Several updates each day**
Tweet celebrities and influential people in the field with personalised messages, do not just copy and paste the URL. Try to vary the updates that you post, sometimes Tweet about a percentage of funds raised and at other times promote specific reward tiers or say what will happen with the funds raised.

**Facebook: Around one update a day**
You can use Facebook milestones occasionally to really catch followers' attention, but try not to overdo this facility and use it only for big milestones like the half-way mark. Images are good to post with the link, but make sure these are the right size. We used neither Facebook ads nor promotion of posts, but would be interested to hear if anyone has had success with these.

**LinkedIn: Two to three updates during the whole campaign.**

**Kickstarter updates: Five spaced throughout the project**
Kickstarter project updates reach people who have already pledged funds to your campaign and people who have chosen to follow your project. They are already interested or motivated by your campaign so these updates should be interesting, friendly and include some sort of call to action, like asking them to share the campaign throughout their own social networks.

### 5. Tell everyone!

Use all of your networks, including the less obvious ones. If you have an interesting, well-designed project, people will want to hear about it, but whatever you do, don't use spam!

Get featured on the blogs and media of your field and work with your institutional press contacts and trusted media. Keep pushing through all of these channels.

### 6. Be flexible

If you have a few hundred backers you have to be able to adapt to their needs. So, if most of your backers are in America and Australia, then your social media should be active at times to likely to reach these people, not only at mid-morning Central European Time.

### 7. The science outreach community won't be your backers' base

Mark Rosin, Guerilla Science, gave us the following advice, and we fully agree:

*Getting other people in the PCST (science outreach) community to try and spread the word is not particularly effective. I'm not sure why, exactly, but my best guess is that everyone is competing for bandwidth (and money), so there's only a limited amount of support one can reasonably expect.*

So, get your message out beyond your network; it is hard but it is necessary.

**Crowdfund your projects, and if you want to...**

...develop a product that you can distribute, **use Kickstarter**
...raise money for a specific activity or event, **use Indiegogo or Globalgiving**
...raise money for a project with scientific interest, **consider RocketHub**
...get some time from an expert, **use 1%Club**

### Acknowledgement

Universe in a Box development was supported by the European Union. We would like to thank Jaya Ramchandani, Erik Arends and Remco van Schadewijk for their contribution to the crowdfunding campaign and comments to this article.

### Further Reading

Kaplan, K. 2013, Nature, 497, 7447

Wheat, R. E. et al. 2013, Trends in Ecology & Evolution, 28, 2

Marlett, D. 2014, *Crowdfunding Art, Science and Technology: A Quick Survey of the Burgeoning New Landscape*. Available at: http://www.academia.edu/6831809/Crowdfunding_Art_Science_and_Technology_A_Quick_Survey_of_the_Burgeoning_New_Landscape

### Links

[1] Kickstarter: https://www.kickstarter.com/

[2] Additional information about Universe in a Box: http://www.unawe.org/resources/universebox/

[3] Campaign video: https://www.youtube.com/watch?v=SVwrYbTiuzg

[4] Campaign Page: https://www.kickstarter.com/projects/unawe/universe-in-a-box

[5] Peerreach, a website used to identify high-impact Twitter users:

Science: http://peerreach.com/lists/science/;

Science UK: http://peerreach.com/lists/science/uk;

Education: http://peerreach.com/lists/education/

| Source | Number of backers | % Pledged | Amount pledged (euros) |
|---|---|---|---|
| Direct traffic (no referrer information) | 97 | 42.46 | 7234 |
| Facebook | 26 | 8.74 | 1545 |
| Twitter | 24 | 9.07 | 1490 |
| universetoday.com | 14 | 5.52 | 941 |
| Other sources | 14 | 15 | 345 |
| Embedded Kickstart widget on www.unawe.org | 9 | 5.05 | 860 |
| Advanced Discovery on Kickstarter | 8 | 3.09 | 526 |
| Google searches | 7 | 14 | 695 |
| unawe.org | 5 | 3.44 | 586 |
| phys.org | 4 | 1.12 | 190 |
| scienceblogs.de | 4 | 0.82 | 140 |
| astronomie.nl | 3 | 1.76 | 300 |
| Leiden University | 3 | 0.56 | 95 |
| allesoversterrenkunde.nl | 2 | 1.17 | 200 |

*Table 2. Overview of the sources for pledges.*

### Biographies

**Abi J. Ashton** is a science communicator with a particular specialisation in delivering hands on science workshops to kids around the world. Since discovering the Universe Awareness educational programme almost a year ago and getting involved, she has worked on developing Universe in a Box and would like to see it used in even more schools around the world.

**Pedro Russo** is the international project manager for the educational programme Universe Awareness. For more information, visit: http://www.unawe.org/russo/

**Thilina Heenatigala** is an astronomy communicator with an emphasis on education. He is affiliated with the Galileo Teacher Training Program as the communication manager and Universe Awareness as the assistant editor of the IAU astroEDU project. You can follow Thilina on www.twitter.com/ThilinaH.